\documentclass[fleqn,usenatbib]{mnras}

\usepackage{newtxtext,newtxmath}
\usepackage[T1]{fontenc}
\usepackage{ae,aecompl}

\usepackage{graphicx}	
\usepackage{amsmath}	
\usepackage{multirow}

\usepackage{tikz}
\usetikzlibrary{arrows,shapes}

\title[Towards solar measurements of nuclear reaction rates]{Towards solar measurements of nuclear reaction rates}
\author[E.\ P.\ Bellinger \& J.\ Christensen-Dalsgaard]{
Earl Patrick Bellinger$^{1,2}$\thanks{E-mail: ebellinger@mpa-garching.mpg.de} and J{\o}rgen Christensen-Dalsgaard$^2$
\\ $^{1}$Max Planck Institute for Astrophysics, Garching, Germany
\\ $^{2}$Stellar Astrophysics Centre, Department of Physics and Astronomy, Aarhus University, Denmark
}
\date{Submitted \today}
\pubyear{2022}

\newcommand{\pp}{\mbox{$\rm{pp}$}}
\newcommand{\pep}{\mbox{$\rm{pep}$}}
\newcommand{\ppfull}{\mbox{$\rm{p}\big(\rm{p}, \rm{e}^+ \nu\big)\rm{d}$}}
\newcommand{\tralphiumalpha}{\mbox{$^3\rm{He}\big(^3\rm{He}, 2\rm{p}\big)\alpha$}}
\newcommand{\hehe}{\mbox{$^3$He--$^3$He}}
\newcommand{\tralphiumbe}{\mbox{$^3\rm{He}\big(\alpha, \gamma\big) {^7\rm{Be}}$}}
\newcommand{\hehef}{\mbox{$^3$He--$\alpha$}}
\newcommand{\BeB}{\mbox{$^7\rm{Be}\big(\rm{p}, \gamma\big) {^8\rm{B}}$}}
\newcommand{\beb}{\mbox{$^7$Be--$\rm{p}$}}
\newcommand{\NO}{\mbox{$^{14}\rm{N}\big(\rm{p}, \gamma\big) {^{15}\rm{O}}$}}

\definecolor{dodgerblue}{rgb}{0.76,0.87,1}

\newif\ifref
\reftrue 
\reffalse
\definecolor{darkred}{rgb}{0.7, 0, 0}
\newcommand{\mb}[1]{\ifref\textcolor{darkred}{#1}\else #1\fi}

\newif\ifreff
\refftrue 
\refffalse
\definecolor{darkred}{rgb}{0.7, 0, 0}
\newcommand{\mbb}[1]{\ifreff\textcolor{darkred}{#1}\else #1\fi}

\newif\ifrefff
\reffftrue 
\reffffalse
\definecolor{darkred}{rgb}{0.7, 0, 0}
\newcommand{\mbbb}[1]{\ifrefff\textcolor{darkred}{#1}\else #1\fi}

\newif\ifreffff
\refffftrue 
\refffffalse
\definecolor{darkred}{rgb}{0.7, 0, 0}
\newcommand{\mbbbb}[1]{\ifreffff\textcolor{darkred}{#1}\else #1\fi}

\newif\ifrefffff
\reffffftrue 
\reffffffalse
\definecolor{darkred}{rgb}{0.7, 0, 0}
\newcommand{\mbbbbb}[1]{\ifrefffff\textcolor{darkred}{#1}\else #1\fi}

\begin{document}
\label{firstpage}
\pagerange{\pageref{firstpage}--\pageref{lastpage}}
\maketitle

\begin{abstract} 
Nuclear reaction rates are a fundamental yet uncertain ingredient in stellar evolution models. 
The astrophysical $S$-factor pertaining to the initial reaction in the proton--proton chain is uncertain at the 1\% level, which contributes a systematic but generally unpropagated error of similar order in the theoretical ages of stars. 
In this work, we study the prospect of improving the measurement of this and other reaction rates in the pp chain and CNO cycle using helioseismology and solar neutrinos. 
We show that when other aspects of the solar model are improved, then it shall be possible using current solar data to improve the precision of this measurement by nearly an order of magnitude, and hence the corresponding uncertainty on the ages of low-mass stars by a similar amount. 
\end{abstract} 


\section{Introduction} 
Just over one hundred years have passed since Arthur Eddington's speculation that stars primarily gain their energy not by contraction, but by tapping into their sub-atomic energy  \citep{1920SciMo..11..297E}. 
As of the work of \citet{1955PASP...67..154H} and \citet{1957ApJ...125..233S} we understand the Sun and other low-mass stars to use the proton--proton chain as their principal energy source, the net reaction of which is the conversion of hydrogen into helium. 
Relying primarily on this process over the past 4.57~Gyr, the Sun has by now exhausted approximately half of its central hydrogen supply. 

The proton--proton chain predominately begins with the fusion of two protons into deuterium, emitting a positron and an electron neutrino: \ppfull{} (hereinafter denoted \pp{}). 
The rate at which the Sun evolves depends on the rate at which this and its other nuclear reactions occur, and hence their cross sections (\mbbbbb{or astrophysical $S$-factors, see e.g., \citealt{2013sse..book.....K}}). 
However, the $S$-factor of the \pp{} reaction is approximately ${4\times 10^{-22}}$~keV~\mbbbbb{barns (b)}, which is too small for modern laboratories to measure. 
Consequently, stellar evolution models rely on a theoretical estimate, which has an uncertainty of approximately 1\% \citep[][]{2011RvMP...83..195A}. 
Theoretical ages of low-mass stars are controlled by this reaction and hence inherit this uncertainty. 


Precise measurements of the solar age, mass, radius, and luminosity make the Sun a promising laboratory for improving measurements of the \pp{} rate and the rates of other nuclear reactions ongoing in the Sun. 
Further constraints come from helioseismic measurements of the Sun's global oscillations, which have revealed the solar interior structure \citep[for reviews, see][]{2016LRSP...13....2B, 2021LRSP...18....2C}, as well as neutrino fluxes from various reactions, which have been measured in detail by several experiments \citep[e.g.,][]{2004IJMPA..19.1167B, 2019PhRvD.100h2004A, 2018arXiv180504163H}. 
These ingredients taken together tightly constrain the space of possible solar evolution models, and hence the possible rates of nuclear reactions. 

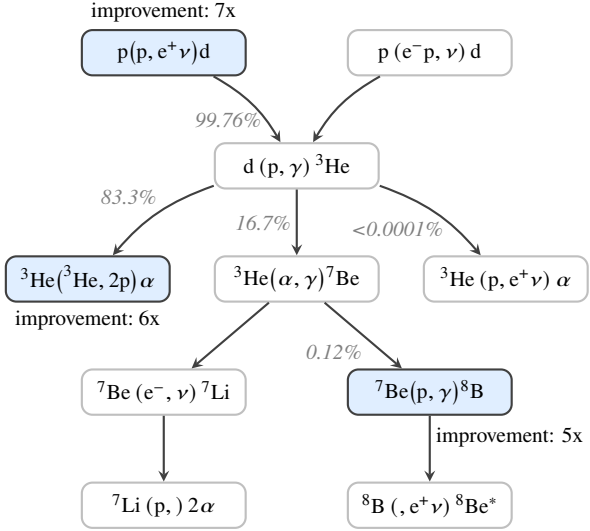
\begin{figure}
    \centering
    \begin{tikzpicture}[bend angle=15]
    
    \tikzstyle{study}=[rounded corners, thick, draw=black!75, fill=dodgerblue!50, 
    minimum size=6mm, text width=7em, text centered]
    \tikzstyle{normal}=[rounded corners, thick, draw=black!25, fill=blue!0, minimum size=6mm, text width=7em, text centered]
    \tikzstyle{edge}=[rectangle,thick,draw=black!75,
  			  minimum size=4mm, -stealth]
    \tikzstyle{lab}=[text centered, text width=1em, minimum size=2mm]
    
        
    	\node [style=study, label=above:{improvement: 7x}]  (0) at (-1.75, 6) {\ppfull{}};
		\node [style=normal] (1) at ( 1.75, 6) {$\rm{p}\left(\rm{e}^-\rm{p}, \nu \right) \rm{d}$};
		\node [style=normal] (2) at ( 0, 4.5) {$\rm{d} \left( \rm{p}, \gamma \right) {^3\rm{He}} $};
		\node [style=study, label=below:{improvement: 6x}]  (3) at (-2.75, 3) {\tralphiumalpha{}};
		\node [style=normal] (4) at ( 0, 3)   {\tralphiumbe{}};
		\node [style=normal] (5) at ( 2.75, 3) { $^3\rm{He} \left( \rm{p}, \rm{e}^+ \nu \right) \alpha $ };
		\node [style=normal] (6) at (-1.75, 1.5) { $^7\rm{Be} \left( \rm{e}^-, \nu \right) {^7\rm{Li}} $ };
		\node [style=normal] (7) at (-1.75, 0) { $^7\rm{Li} \left(\rm{p}, \right) 2\alpha $ };
		\node [style=study]  (8) at ( 1.75, 1.5) {\BeB{}};
		\node [text width=6.5em] (10) at (2.75, 0.9) {improvement:~5x};
		\node [style=normal] (9) at ( 1.75, 0) { $ ^8\rm{B} \left(, \rm{e}^+ \nu \right) {^8\rm{Be}^\ast} $ };
		
		\draw [style=edge, bend left ] (0) to node[text width=1.5cm,below,align=left] {$\textit{\textcolor{gray}{99.76\%}}$} (2);
		\draw [style=edge, bend right] (1) to (2);
		\draw [style=edge, bend right] (2) to node[text width=1.5cm,above,align=left] {$\textit{\textcolor{gray}{83.3\%}}$} (3);
		\draw [style=edge] (2) to node[text width=0.7cm, left,align=left] {$\textit{\textcolor{gray}{16.7\%}}$} (4);
		\draw [style=edge, bend left ] (2) to node[text width=2.2cm, below,align=left] {$\textit{\textcolor{gray}{<0.0001\%}}$} (5);
		\draw [style=edge]             (4) to (6);
		\draw [style=edge]             (4) to node[text width=1.5cm, below,align=left] {$\textit{\textcolor{gray}{0.12\%}}$} (8);
		\draw [style=edge]             (6) to (7);
		\draw [style=edge]             (8) to (9);
	
	
    \end{tikzpicture}   
    \caption{\mbb{The proton--proton chain, with branching ratios indicated. We show in this work that the Sun can in principle be used as a laboratory to improve current uncertainties on estimates of the \pp{}, \hehe{}, and \beb{} reaction rates by factors of 7, 6, and 5, respectively.} \label{fig:ppchaincno} }
\end{figure}

\mb{Using 6 months of helioseismic data, \citet{2000BASI...28..105A} constrained the \pp{} rate to 6\%. In this work, we aim to show that by using the full data set, in combination with solar flux measurements, this constraint can in principle be improved to 0.14\%. 
Similarly, solar data can be used to improve the current level of uncertainty on both the \tralphiumalpha{} and \BeB{} rates by a factor of five or more.} \mbb{The pp chain and a summary of some of the findings of this paper are shown in Figure~\ref{fig:ppchaincno}.}

Helioseismic measurements have shown that a standard stellar evolution model at the solar age matches the internal density profile of the Sun within 1\%. 
While significant differences do remain between the model and the observations, this level of agreement instils some confidence that the standard model of the Sun is essentially right, and points toward small adjustments in the microphysics used to construct the model as being the cause of the discrepancies. 
In particular, these discrepancies lie mainly at the base of the solar convection zone, which is suggestive that the main improvements needed to the solar model are needed in the calculations of radiative opacities \citep{2017Atoms...5...22.}. 
\mb{Additionally, longstanding problems still remain regarding the abundances of the solar atmosphere \citep[e.g.,][]{2014dapb.book..245B} though a solution has very recently been claimed \citep{2022arXiv220302255M}.}

Under the assumption that the remaining problems with the standard solar model can be resolved, we aim here to quantify the level of improvement that current measurements of the solar data can provide to measurements of nuclear reaction rates. 
\mbbbb{We will first assume a standard solar model to be generally correct in its construction. 
Then, we determine how well these rates can be constrained via stellar evolution theory using the present uncertainties in the solar data: the solar age, radius, luminosity, metallicity, observed neutrino fluxes, and low-degree helioseismic observations}. 
\begin{figure*}
    \centering
    \includegraphics[width=0.6\textwidth, keepaspectratio, trim={0.5cm 0.5cm 0.5cm 0.5cm}, clip]{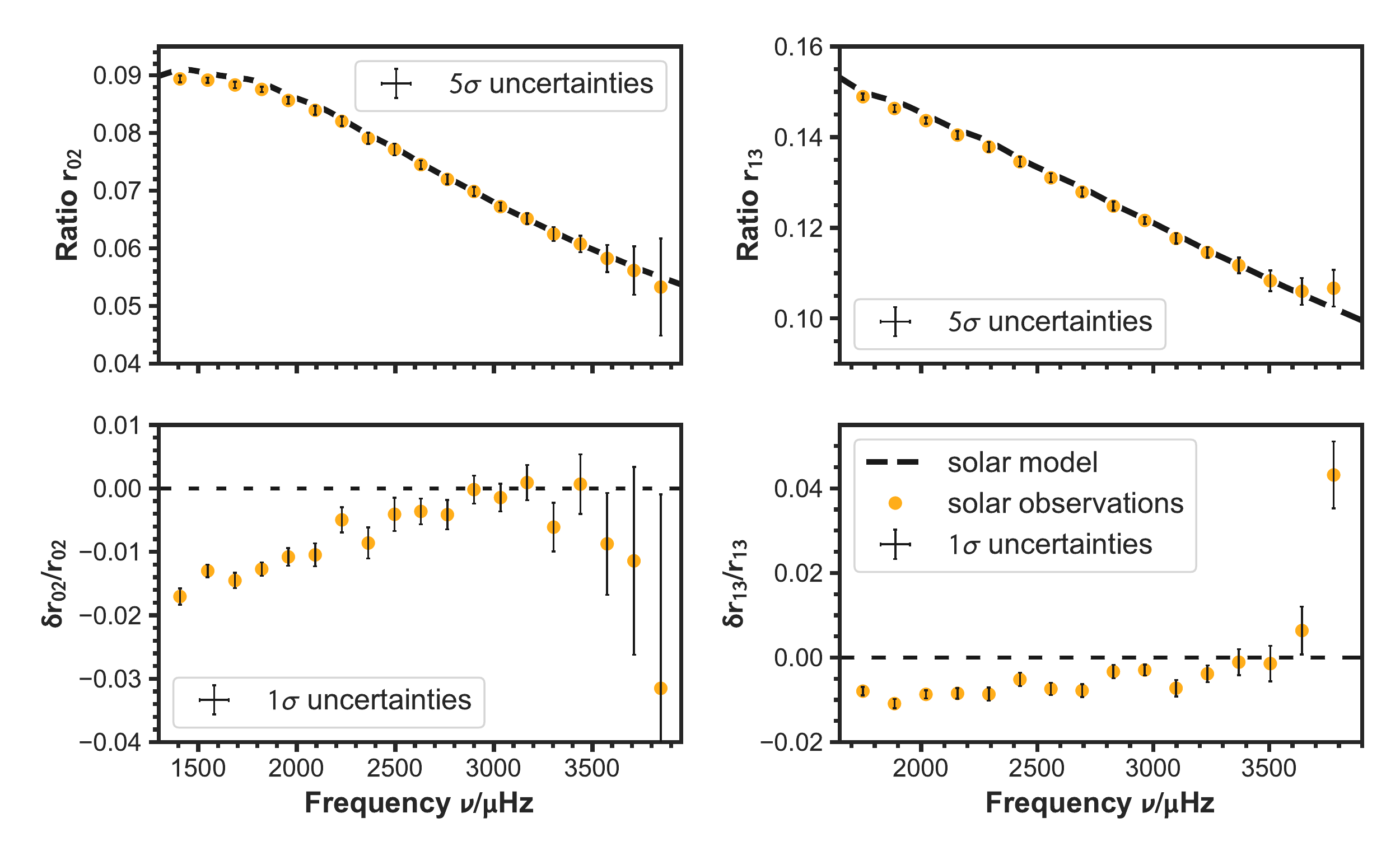}%
    \includegraphics[width=0.4\textwidth, keepaspectratio, trim={0.5cm -0.95cm 0.5cm 0.5cm}, clip]{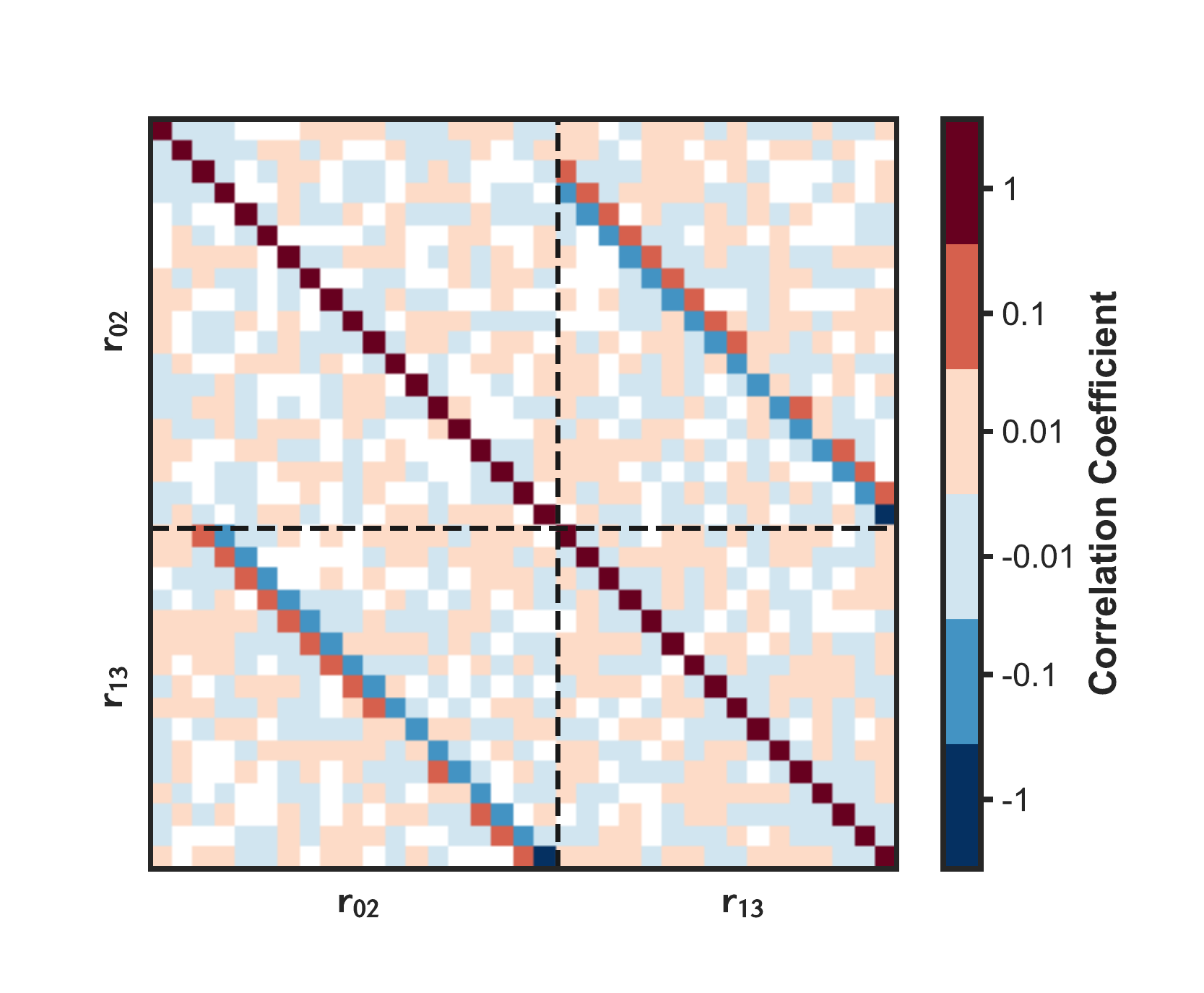}\\%
    \caption{Frequency separation ratios $r_{02}$ and $r_{13}$ (Equation~\ref{eq:ratios}), observed from helioseismology and computed from a standard solar model. The top panels show the ratios themselves, and the bottom panels show the relative differences in the ratios, here defined as $($observed$-$model$)/$model. \mbbbb{The rightmost panel visualises the correlation matrix of the measured frequency separation ratios shown in the left panels, ordered also in increasing frequency.} }
    \label{fig:ratios}
\end{figure*}

Improved measurements of nuclear reaction rates are useful for improving the predictions of stellar evolution theory, which are important for related fields that rely on accurate stellar ages, such as galactic archaeology \citep[e.g.,][]{2017AN....338..644M}. 
Additionally, improved measurement of the \pp{} rate in particular has the potential to drive down uncertainties in electroweak theory, which underpins its current estimate.  \mbbbb{\citet{2016PhLB..760..584A} quantified the theoretical uncertainties of \pp{} fusion in the context of chiral effective field theory and found that the dominant contribution stemmed from the uncertainties in low-energy constants, i.e., the parameters of the nuclear interaction model, which they obtained from the analysis of \citet{2016PhRvX...6a1019C}. 
These parameters are constrained by fits to low-energy experiments, and can in principle be predicted by the theory of quantum chromodynamics \citep{2015PhRvL.114e2501B}. 
A precise measurement of the \pp{} rate from the Sun may then be useful in driving down the uncertainties in these low-energy constants and hence improve our understanding of the relevant theories. } 
A conflict between the predictions of these theories and stellar evolution would furthermore be indicative of underlying issues or missing physics in one or the other \citep[such as non-standard interactions, e.g.,][]{2021JCAP...07..042S}, as was the case with the solar neutrino problem and neutrino oscillations \citep{1990PhRvL..65.2233B}, and hence also potentially drive progress. 

The paper is organised as follows. Section~\ref{sec:SSM} describes the construction of our solar models. Section~\ref{sec:data} gives the observational constraints from helioseismology. Section~\ref{sec:rates} shows the effects of changing various nuclear reaction rates on the solar structure and resulting observations, and derives simple lower bounds on the possible solar constraints on the rates. Section~\ref{sec:fluxes} performs a similar analysis using the neutrino flux measurements. Section~\ref{sec:MCMC} presents a full Markov chain Monte Carlo (MCMC) analysis of all known sources of uncertainty to the solar model and provides more accurate estimates on the possible constraints. Finally, Section~\ref{sec:ages} explores the implications for stellar ages, and Section~\ref{sec:conclusion} summarises the work.







\section{Standard solar models} \label{sec:SSM}
In this work we make use of standard solar models \citep[SSMs, see e.g.,][]{1972ARA&A..10...25B}. 
These are $1~\rm{M}_\odot$ stellar evolution models defined as having the solar radius, solar luminosity, and solar \mbbbbb{surface} metallicity at the solar age. 
To achieve this, we use the Aarhus Stellar Evolution Code \citep[\textsc{Astec},][]{2008Ap&SS.316...13C} and calibrate the convective mixing length parameter $\alpha_{\text{MLT}}$ as well as the initial helium abundance $Y_0$ and initial metallicity $Z_0$ using numerical optimisation. 

\mbbbbb{The calculations use the OPAL equation of state \citep{1996ApJ...456..902R, 2002ApJ...576.1064R}, OPAL opacities \citep{1996ApJ...464..943I} at temperatures above $10^4 \,{\rm K}$ and low-temperature opacities from \citet{2005ApJ...623..585F}. Nuclear reaction rates were obtained from \citet{2011RvMP...83..195A}, assuming electron weak screening \citep{1954AuJPh...7..373S}. The atmospheric structure is represented by a relation between optical depth and temperature approximating Model C of \citet{1981ApJS...45..635V}.}

\mbbbbb{We include the effects of element diffusion, as approximated by \citet{1993ASPC...44..439M}, representing heavy elements by fully ionized oxygen. Further details on solar modelling are provided by \citet{2021LRSP...18....2C}.}

\mbbbbb{We assume the GS98 solar composition \citep{1998SSRv...85..161G} as it currently yields the solar model in closest agreement with helioseismic inferences, although as our goal is a differential analysis, the specific solar composition should have little effect. Similarly, the exact values of the assumed global solar properties are not important for the analysis.}
That being said, we assume the following solar parameters:
\begingroup
\allowdisplaybreaks
\begin{align*} \label{eq:solar-vals}
    \text{mass } \rm{M}_\odot &= 1.989 \times 10^{33}\; \text{g} \\
    \text{radius } \rm{R}_\odot &= 6.9599 \times 10^{10}\; \text{cm} \\
    \text{luminosity } \rm{L}_\odot &= 3.828 \times 10^{33}\; \text{erg}\; \text{s}^{-1} \\
    \text{heavy mass fraction } (Z/X)_\odot &= 0.02307 \\
    \text{age } \tau_\odot &= 4.57\times 10^{9} \; \text{yr}.
\end{align*}
\endgroup
For our initial analysis, we shall regard these values as being fixed; in Section~\ref{sec:MCMC}, however, we propagate their uncertainty as well. 
The calibration using these values results in a SSM with ${Y_0 = 0.2754}$, ${Z_0 = 0.01847}$, and ${\alpha_{\text{MLT}} = 1.9118}$.

\section{Helioseismic data} \label{sec:data}
We have obtained low-degree solar oscillation data from the Birmingham Solar Oscillations Network \citep[BiSON,][]{2014MNRAS.439.2025D}. From these measurements we calculate the frequency separation ratios of \citet{2003A&A...411..215R}, which probe the interior conditions of the Sun \mbbbbb{\citep{2005MNRAS.356..671O}} and facilitate a comparison of theory and observation:
\begin{equation} \label{eq:ratios}
    r_{02}(n) = \dfrac{\nu_{n, 0} - \nu_{n-1, 2}}{\nu_{n,1} - \nu_{n-1, 1}}, \qquad
    r_{13}(n) = \dfrac{\nu_{n, 1} - \nu_{n-1, 3}}{\nu_{n+1,0} - \nu_{n, 0}}. 
\end{equation}
Here $\nu_{n,\ell}$ refers to the frequency of the mode with radial order~$n$, which in the observations ranges from 9 to 27, and spherical degree~$\ell$, which ranges from 0 to 3. 
As is evident by these equations, the ratios are correlated; here we calculate the covariance matrix via 100,000 Monte Carlo iterations. 

We have used the Aarhus adiabatic oscillation package \citep[\textsc{Adipls},][]{2008Ap&SS.316..113C} to calculate the theoretical frequencies of our models. A comparison of the observed helioseismic frequency separation ratios to the theoretical frequency separation ratios of a SSM are shown in Figure~\ref{fig:ratios}. 
Qualitatively, the match is good. As is well-known, however, there is significant tension between the observed and computed ratios. Indeed, the ratios differ by up to $14\sigma$, with the largest differences being at low frequency. 
This arises because, as mentioned, while the internal structure of a standard solar model is close to that of the Sun, they are not exactly the same. 

\section{Nuclear reaction rates} \label{sec:rates}
In this work, we consider the variation of five nuclear reaction rates, which are tabulated along with their astrophysical $S$-factors in Table~\ref{tab:Sfactors}. 
For an initial analysis, we have considered the effect of changing each rate in isolation by factors of its standard deviation~$\sigma$. 
We then calibrate a new SSM corresponding to each value. 

\begin{table}
    \centering
    \caption{Astrophysical $S$-factors of Solar Nuclear Reactions \label{tab:Sfactors}}
    \begin{tabular}{l l l l}\hline
        \multirow{2}{*}{Reaction} & \hphantom{[}$S(0)$ & \multicolumn{2}{c}{Uncertainty} \\
        & [keV b] & [keV b] & [\%] \\\hline\hline
        \hphantom{$^{14}$}\ppfull{} & $4.01 \times 10^{-22}$ & $0.04 \times 10^{-22}$ & $1.0\%$ \\
        \hphantom{$^{1}$}\tralphiumalpha{} & $5.21 \times 10^3$ & $0.27 \times 10^3$ & $5.2\%$ \\
        \hphantom{$^{1}$}\tralphiumbe{} & $0.56$ & $0.03$ & $5.4\%$ \\
        \hphantom{$^{1}$}\BeB{} & $2.08 \times 10^{-2}$ & $0.16 \times 10^{-2}$ & $7.7\%$ \\
        \NO{} & $1.66$ & $0.12$ & $7.2\%$ \\
        \hline
    \end{tabular}\\[0.5\baselineskip]
    \textbf{Note.} $S$-factors and uncertainties obtained from \citealt{2011RvMP...83..195A}. 
\end{table}

\begin{figure}
    \centering
    \includegraphics[height=8.7cm,keepaspectratio]{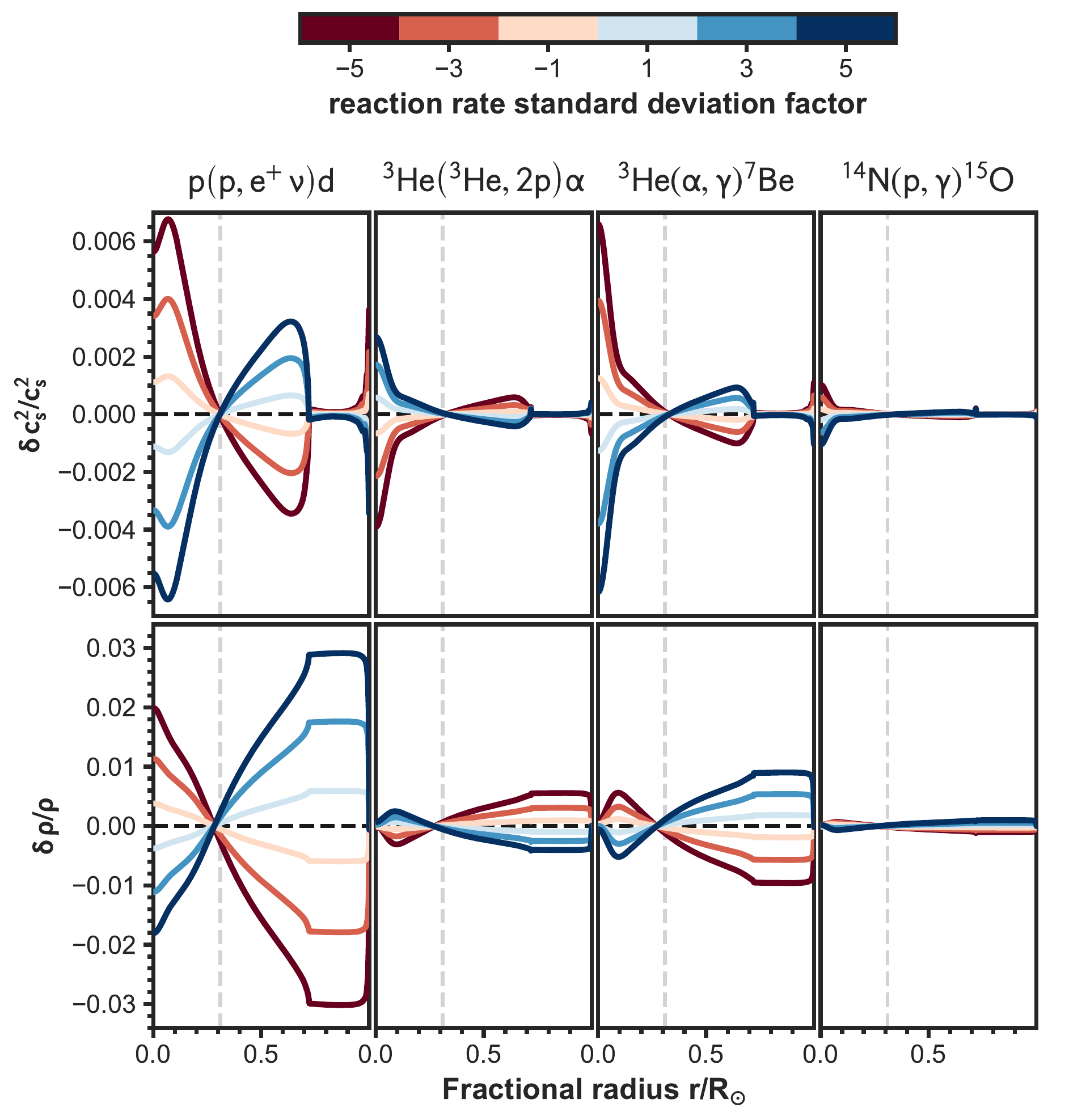}
    \caption{The effects of modifying nuclear reaction rates (listed atop each panel) on the squared adiabatic sound speed profile (top panels) and the density profile (bottom panels) of a standard solar model. Each line corresponds to a standard solar model, with the relevant rate of that model being multiplied by some factor of its standard deviation (\emph{cf}.~Table~\ref{tab:Sfactors}) as shown in the colourbar. The dashed vertical line shows where the sound speed differences change sign. \label{fig:rates-struc}}
\end{figure}

Figure~\ref{fig:rates-struc} shows the corresponding changes to the internal structure of the SSM in terms of the differences in density $\rho$ and squared adiabatic sound speed $c_s^2$. 
\mbbbb{These are the most relevant quantities to inspect because they determine the solar oscillation spectrum. 
The sound speed is essentially determined by the balance of the temperature and the mean molecular weight.} 
\mbbbbb{Unlike the other reactions, modifications to the \pp{} reaction result in differences to the mean molecular weight that peak away from the centrepoint, which causes the sound speed differences to behave similarly. On the other hand, the differences to the temperature change sign at a fractional radius of $\sim 0.3$ across all of the reactions, approximately corresponding to the point beyond which nuclear reactions become very inefficient, thus explaining why the differences across the various reactions pass through zero at essentially the same point. } 
Overall, the largest differences to the solar structure are caused by changes to the \pp{} rate, in which an increase of 1$\sigma$ causes a change of 0.5\% to the central solar density. 
\mbbbb{In order to maintain the solar parameters, increasing the \pp{} rate by 1$\sigma$ requires a 0.05\% decrease to the protosolar metallicity, a 0.03\% decrease to the protosolar hydrogen abundance, and a 0.2\% increase to the mixing length parameter. Changes to the other rates result in smaller differences because, despite their larger uncertainty, the resulting changes to the solar structure are smaller. } 

As is well-known, the largest discrepancy between the solar structure and SSMs is at the base of the convection zone \citep[e.g.,][]{2009ApJ...699.1403B}. In all of the cases investigated here, the greatest changes to the sound speed are near to the solar centre. Therefore, inaccurate nuclear reaction rates cannot be the dominant explanation of the discrepancies between the Sun and standard models. 

\begin{figure}
    \centering
    \includegraphics[height=8.7cm,keepaspectratio]{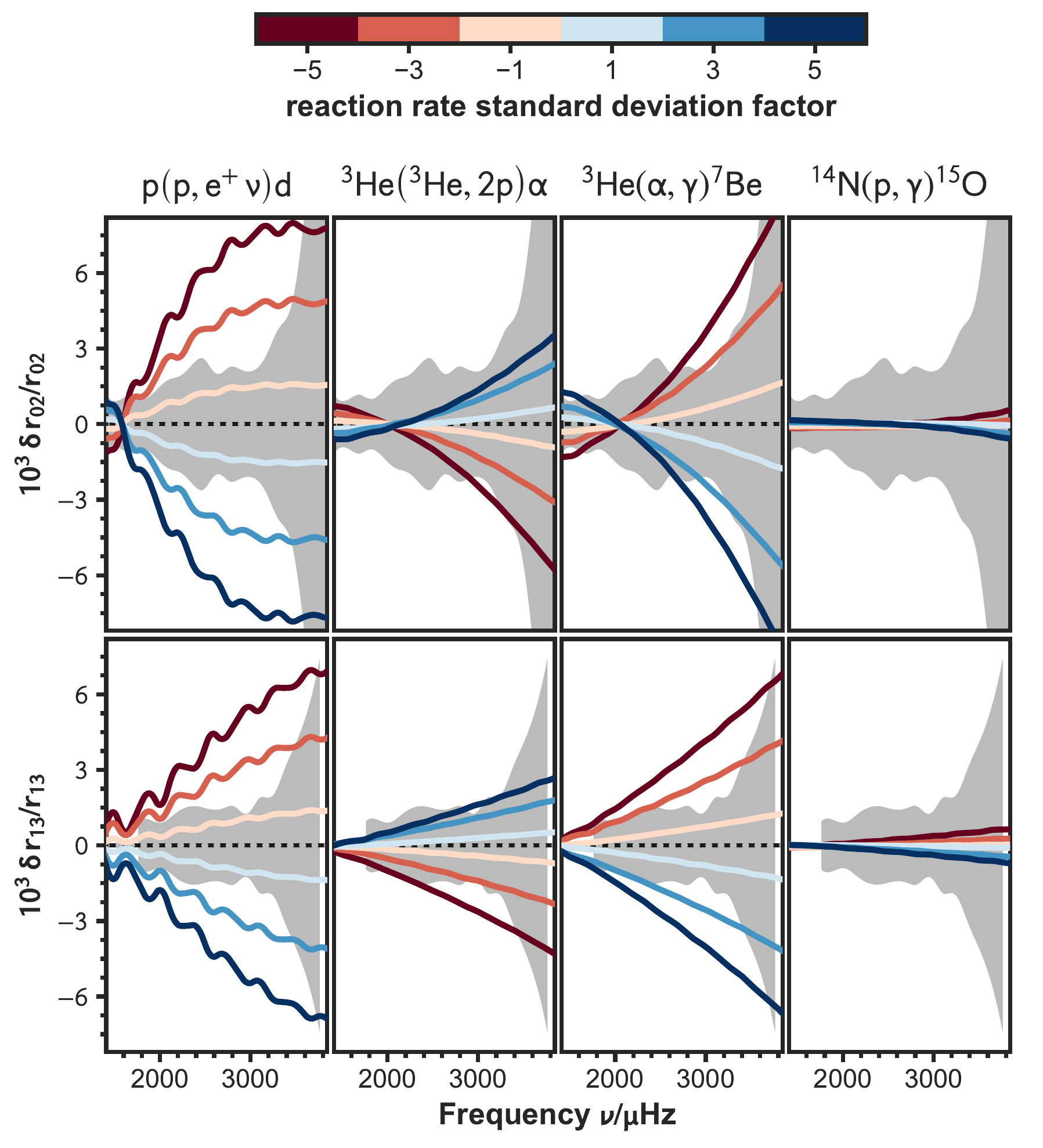}
    \caption{The effects of modifying nuclear reaction rates (listed atop each panel) on the frequency separation ratios $r_{02}$ (top panels) and $r_{13}$ (bottom panels) of a standard solar model. Each line corresponds to a recalibrated standard solar model, with the relevant rate being multiplied by some factor of its standard deviation (\emph{cf}.~Table~\ref{tab:Sfactors}) as shown in the colorbar. The shaded background shows the uncertainties in the helioseismic ratios (\emph{cf}.~Figure~\ref{fig:ratios}). \label{fig:rates}}
\end{figure}

Now we turn our attention to the helioseismic effects of these changes. 
Figure~\ref{fig:rates} shows the changes to the frequency separation ratios due to changes in nuclear reaction rates. 
In this figure it can be seen that increasing the \pp{} rate by $1\sigma$ decreases each of the ratios by $1\sigma$, which implies that a strong constraint can be obtained on this rate from helioseismology. 
\mbbbb{Here it is also evident that changing the rates of the \hehe{} and \hehef{} reactions have effects of similar magnitude, but opposite sign on the helioseismic observations, likely owing to their opposing effect on the pp~chain}. 
\mbbb{Since changes to the \hehef{} reaction imparts the largest change the central sound speed, improved precision to the measurements of high-frequency modes will especially help constrain its rate.}

\begin{figure}
    \centering
    \includegraphics[width=\columnwidth, keepaspectratio]{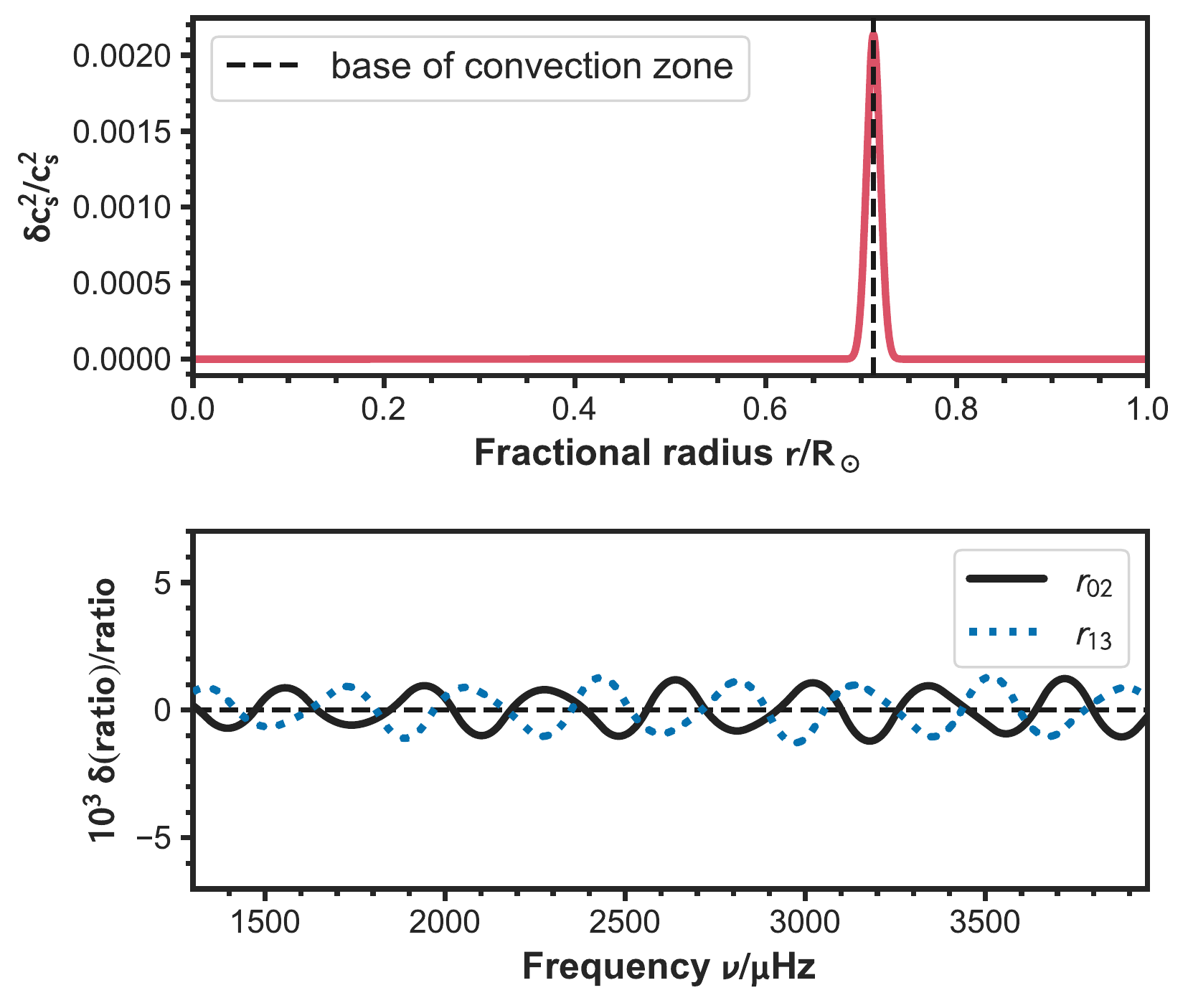}
    \caption{\mbbb{Artificial perturbation to the sound speed profile at the base of the convection zone (top panel) and resulting perturbations to the frequency separation ratios computed through the structure kernels (bottom panel).}}
    \label{fig:bcz}
\end{figure}

\mbbb{
Figure~\ref{fig:rates} also reveals an oscillation in the changes to the ratios due to changes in the \pp{} rate, which gives it a potentially unique signature. 
This change can be attributed to a difference in the sound speed at the base of the convection zone. 
\mbbbbb{While changes to any of the rates perturb the structure there, the magnitude of the effect from the \pp{} rate is several times larger than from any of the others.} 
In order to confirm this as the source of the oscillatory behaviour, we perform a linear perturbation analysis \citep{1994A&A...283..247M, 1994MNRAS.268..880R}. 
The difference in an oscillation mode frequency $\nu$ between a pair of solar models can be estimated through the differences in their internal sound speed and density profiles: 
\begin{equation}
    \dfrac{\delta \nu}{\nu}
    =
    \int_0^R
    K^{(c^2, \rho)}
    \dfrac{\delta c^2}{c^2}
    +
    K^{(\rho, c^2)}
    \dfrac{\delta \rho}{\rho}
    \; \text{d}r
\end{equation}
where $K$ are kernels obtained through the linear perturbation analysis \citep[e.g.,][]{1991sia..book..519G}. 
We introduce a small perturbation to the sound speed at the base of the convection zone and calculate resulting differences the frequency separation ratios using this equation, assuming no change to the density. 
The result, shown in Figure~\ref{fig:bcz}, is an oscillation \mbbbbb{corresponding to the shift with changing frequency in the phase of the eigenfunctions at the base of the convection zone} in the ratio differences with approximately the right periodicity and \mbbbb{within an order of magnitude of the right amplitude.} 
}

\subsection{Reaction rate uncertainties from helioseismology}
We now seek to determine constraints on the nuclear reaction rates that would be possible with current data if the SSM were otherwise correct in its construction. 
We do this by comparing each of the SSMs calibrated with different nuclear reaction rates to the SSM with the nominal values, which we call the reference model. 
As these changes are here considered in isolation, and only consider the random errors in the measurements, they represent the most optimistic case. 

We define the goodness-of-fit metric $\chi^2$ to be
\begin{equation}
    \chi^2
    = 
    \mathbf{R}^{\text{T}} \boldsymbol{\Sigma}^{-1} \mathbf{R}
    ,\qquad 
    R_i 
    = 
    D_i - M_i. \label{eq:chi2}
\end{equation}
Here $\mathbf D$ are the frequency separation ratios (\emph{cf}.~Eqn.~\ref{eq:ratios}) of the reference model, $\mathbf M$ are the ratios of a SSM with changed nuclear reaction rates, and $\mathbf \Sigma$ is the variance--covariance matrix of the measured frequency separation ratios (visualised in Figure~\ref{fig:ratios}). The $1\sigma$ uncertainty is obtained when ${\chi^2=1}$. 

\begin{figure}
    \centering
    \includegraphics[width=\columnwidth, trim={0.5cm 0.5cm 0.5cm 0.1cm}, clip]{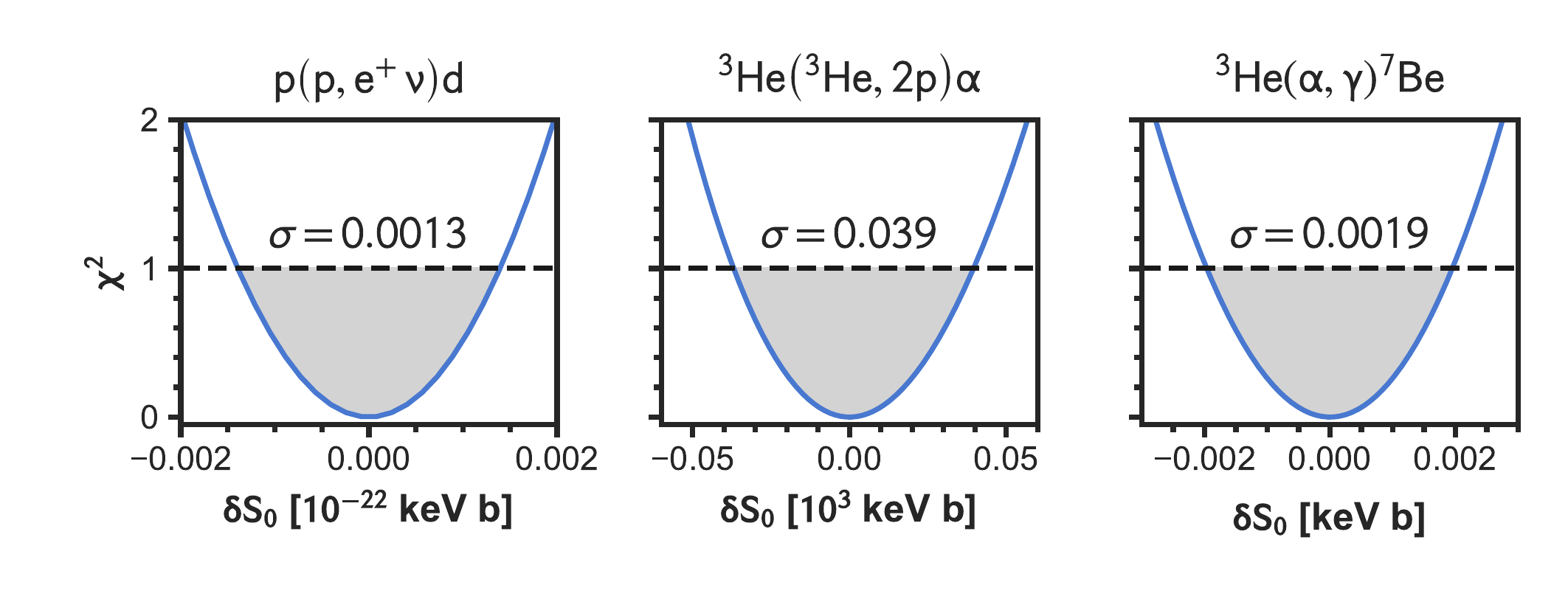}%
    \caption{The agreement with helioseismology as a function of differences in the astrophysical $S$-factors for three nuclear reactions, made under the assumption that the reference model is otherwise correct. 
    The shaded region shows ${\chi^2<1}$.
    The width at ${\chi^2=1}$ gives the theoretical minimum measurement uncertainty $\sigma$ that would be possible from present helioseismic observations.
    These uncertainties reduce the Adelberger uncertainties listed in Table~\ref{tab:Sfactors} by factors of \mbbbbb{30, 7, and 16}, respectively.
    Helioseismology offers comparatively little \mbbbbb{improvement} to the other rates considered and so are not shown. } 
    \label{fig:chi2s}
\end{figure}

Figure~\ref{fig:chi2s} shows the comparisons of the models and furthermore the theoretical improvement from helioseismology to the $S$-factors of the \pp{}, \hehe{}, and \hehef{} reactions. The other two reactions listed in Table~\ref{tab:Sfactors} are not improved with current helioseismic data. 
It can be seen in this figure that helioseismology offers \mbbbbb{more than an order-of-magnitude improvement to the \pp{} and \hehef{} rates as well as a factor seven improvement to the \hehe{} rate}. 

\section{Neutrino fluxes} \label{sec:fluxes}
We shall now perform a similar analysis as in the preceding sections, but here making use of neutrino flux measurements. 
We have obtained the observed solar neutrino fluxes after correction for flavour mixing from \citet{2016JHEP...03..132B} and tabulated the relevant values in Table~\ref{tab:fluxes}. We have also calculated the corresponding fluxes from our reference SSM by integrating the neutrino flux per unit mass over the mass of the model. Note that while the \pp{}, \pep{}, and $^7$Be fluxes are in good agreement with the SSM, the flux from $^8$B has tension exceeding $4\sigma$. 

\begin{figure}
    \centering
    \includegraphics[width=\columnwidth]{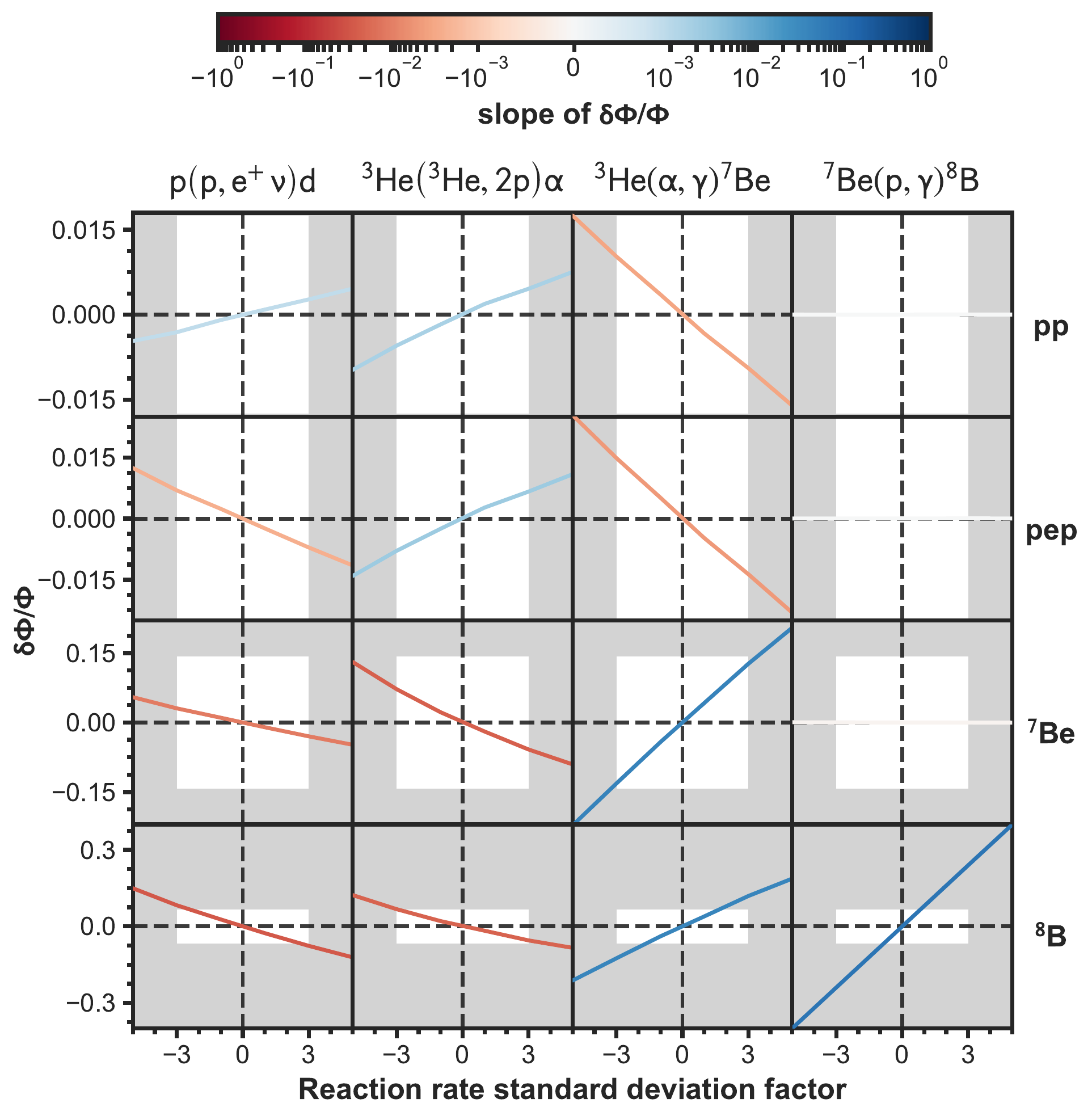}%
    \caption{Changes to the SSM neutrino fluxes (rows) due to changes in nuclear reaction rates (columns). The horizontal axis shows changes to the $S$-factor of each reaction, listed atop each panel, in terms of its standard deviation as listed in Table~\ref{tab:Sfactors}. The vertical axis shows the corresponding change to the flux $\Phi$ from each source, listed on the right. The results are coloured by the slope of a linear fit as given in the colourbar, with blue indicating that increasing the $S$-factor increases the neutrino flux, and red indicating the opposite. \mbb{The grey background shades regions of 3 standard deviations and higher, indicating regions that are ruled out by the observations: the absicssa corresponds to the uncertainties in the reaction rates (as provided in Table~\ref{tab:Sfactors}) and the ordinate corresponds to the uncertainties in the observed fluxes (Table~\ref{tab:fluxes})}.} 
    \label{fig:fluxes}
\end{figure}

\begin{figure}
    \centering
    \includegraphics[width=\columnwidth, trim={0.5cm 0.5cm 0.5cm 0.1cm}, clip]{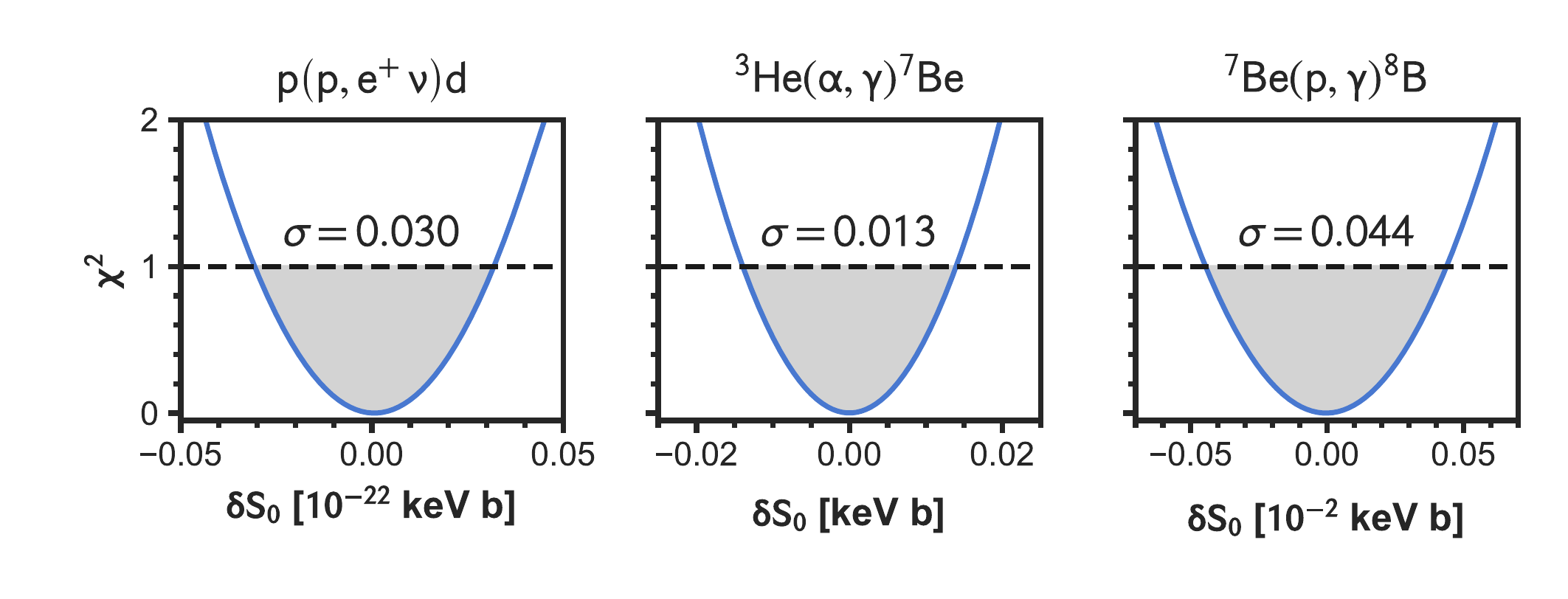}%
    \caption{The agreement with neutrino flux measurements as a function of differences in the astrophysical $S$-factors for three nuclear reactions, made under the assumption that the reference model is otherwise correct.
    These uncertainties reduce those listed in Table~\ref{tab:Sfactors} by factors of 1.3, 2.3, and 3.6, respectively.} 
    \label{fig:chi2s-neutrinos}
\end{figure}

Figure~\ref{fig:fluxes} shows the change in neutrino fluxes from changes to the nuclear reaction rates. The best constraint is provided by $^8$B as its flux is significantly impacted by all of the shown reactions. 
For example, a $1\sigma$ change to the \beb{} reaction rate produces a $3\sigma$ change to the resulting $^8$B flux. 
On the other hand, the \pp{} and \pep{} fluxes do not constrain the reaction rates at all. This figure also demonstrates that, unlike helioseismic data, neutrino fluxes can be used to distinguish changes between \hehe{} and \hehef{}. 
Lastly, from the uncertainties shown in this figure, it is again evident that a constraint on nuclear reaction rates is possible with current solar neutrino flux data. 
This is illustrated in Figure~\ref{fig:chi2s-neutrinos}, which shows the constraints to the solar nuclear reaction rates given by neutrino flux measurements. These measurements clearly have the potential to reduce the uncertainty in the the \pp{}, \hehef{}, and \beb{} reactions, in the last case with an improvement by a factor of $3.6$. 

\begin{table}
    \centering
    \caption{Neutrino Fluxes from Solar Nuclear Reactions \label{tab:fluxes}}
    \begin{tabular}{l l l l l}\hline
         & \multicolumn{3}{c}{Flux $\Phi$ [$10^{10}$~cm$^{-2}$~s$^{-1}$]} \\
        Source & SSM & Measured & Uncertainty & Tension \\ \hline\hline
        \pp{} & 5.977 & 5.971 & 0.035 & $0.2\sigma$ \\
        \pep{} & 1.464 $\times~10^{-2}$ & 1.448 $\times~10^{-2}$ & 0.013  $\times~10^{-2}$ & $1.2\sigma$ \\
        $^7$Be & 0.511 & 0.480 & 0.023 & $1.3\sigma$ \\
        $^8$B & 5.61 $\times~10^{-4}$ & 5.16 $\times~10^{-4}$ & 0.11 $\times~10^{-4}$ & $4.1\sigma$ \\
        \hline
    \end{tabular}\\[0.5\baselineskip]
    \textbf{Note.} Measurements obtained from \citealt{2016JHEP...03..132B}. 
\end{table}

\section{Full error propagation with MCMC} \label{sec:MCMC}
In the previous sections, we explored how each nuclear reaction rate affects the resulting helioseismic data and neutrino fluxes. We furthermore used these relations to derive optimistic lower bounds on the constraints that can be placed on the astrophysical $S$-factors of each rate. 

In this section, we aim to propagate all uncertainties in the solar data, and combine the helioseismic and neutrino flux data in order to obtain realistic uncertainties on the $S$-factors. 
This is similar to the analysis of \citet{2018MNRAS.477.1397S}, except that we shall here assume the SSM to be correct, in order to determine how precisely solar data will be able to measure nuclear reaction rates. 

To that end, we use Markov chain Monte Carlo \citep[MCMC,][]{2010CAMCS...5...65G} to estimate the posterior distributions of the nuclear reaction rates. For priors on the input solar data, we assume the following uncertainties, which are discussed in more detail in the review on solar structure and evolution of \citet{2021LRSP...18....2C}. From \citet{1995RvMP...67..781B} we adopt an age uncertainty of $0.006$~Gyr. 
From a comparison of the \citet{2009ARA&A..47..481A} abundances with meteoric data, we adopt an [Fe/H] uncertainty of $0.04$. From \citet{2008ApJ...675L..53H} we adopt an uncertainty on the solar radius of $0.00002$~R$_\odot$. From the variation in solar irradiance over a solar cycle \citep{2004A&ARv..12..273F}, we adopt a luminosity uncertainty of $0.001$~L$_\odot$.  

As nearly everywhere in the calculations the solar mass appears along with the gravitational constant, the product of which is extremely well known, we keep both quantities fixed at their central values. 
The resulting estimate will therefore be optimistic, but only very slightly so. 
\mbb{We also do not consider uncertainty pertaining to the equation of state, opacities, or the ratio of heavy mass elements. The latter is, however, partially considered through the adopted uncertainty on [Fe/H].} 

The likelihood function we adopt for the analysis is again Eqn.~\ref{eq:chi2}, only extended to include the differences in the neutrino flux data, the solar radius, luminosity, and [Fe/H]. \mbbbbb{We used the \texttt{emcee} \citep{2013PASP..125..306F} package in Python and ran the analysis for 350,000 iterations.} 

\subsection{Results}
The results of the MCMC analysis are presented in Table~\ref{tab:MCMC}. It can be seen that an improved solar model will be able to reduce the uncertainty on the astrophysical $S$-factor of the \pp{} rate by a factor of 7, of the \hehe{} rate by a factor of 6, and the \beb{} rate by a factor of 5. 

\begin{figure}
    \centering
    \includegraphics[width=\columnwidth, trim={0cm 0cm 6cm 6cm}, clip]{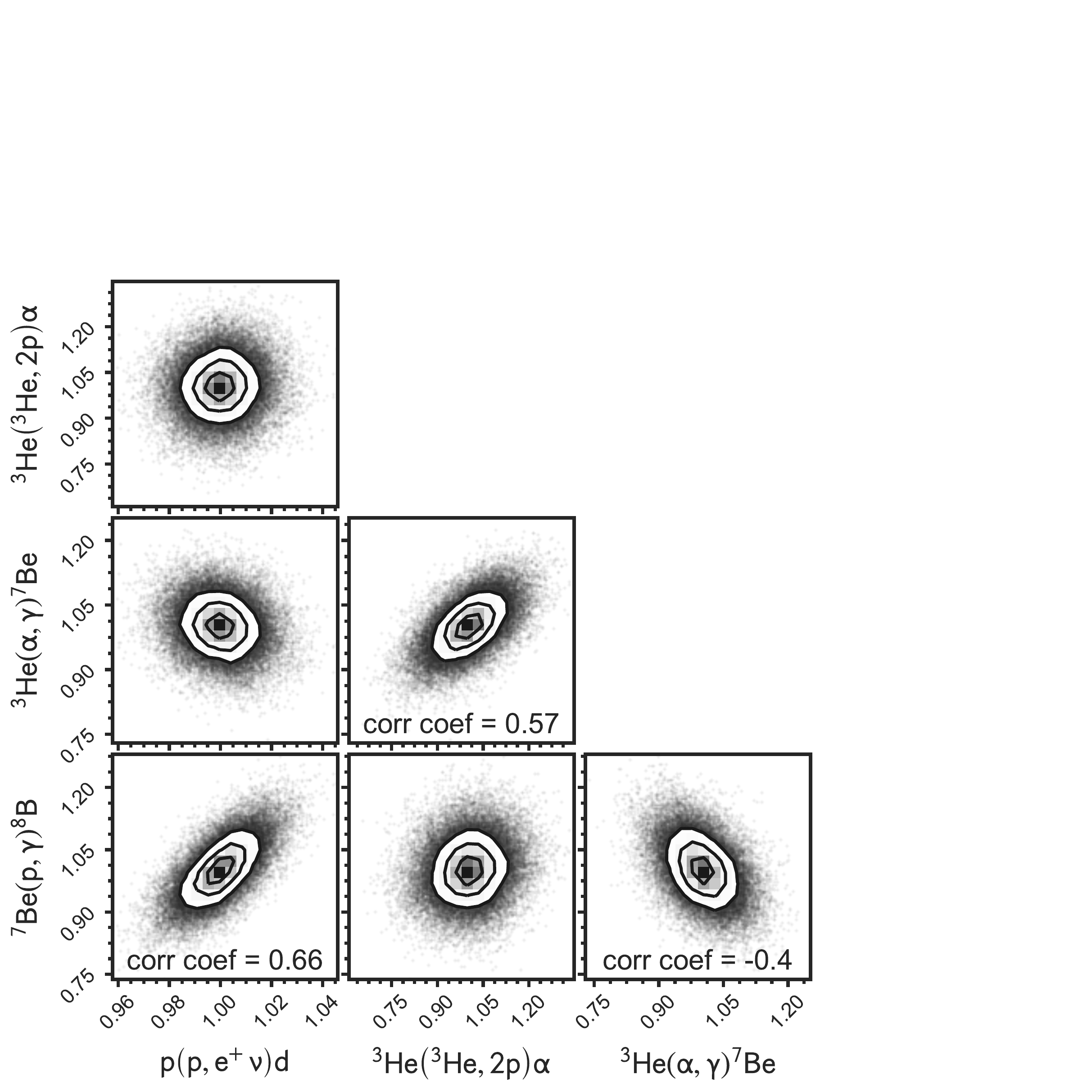}%
    \caption{Scatterplot matrix showing the MCMC results for various pairs of nuclear reaction rates as constrained by solar data. Significant correlations are indicated in the panels. Note that the values on the axes have been normalised such that the mean value is 1. } 
    \label{fig:correlations}
\end{figure}

\begin{figure}
    \centering
    \includegraphics[width=\columnwidth, trim={2cm 0.5cm 0.8cm 0.6cm}, clip]{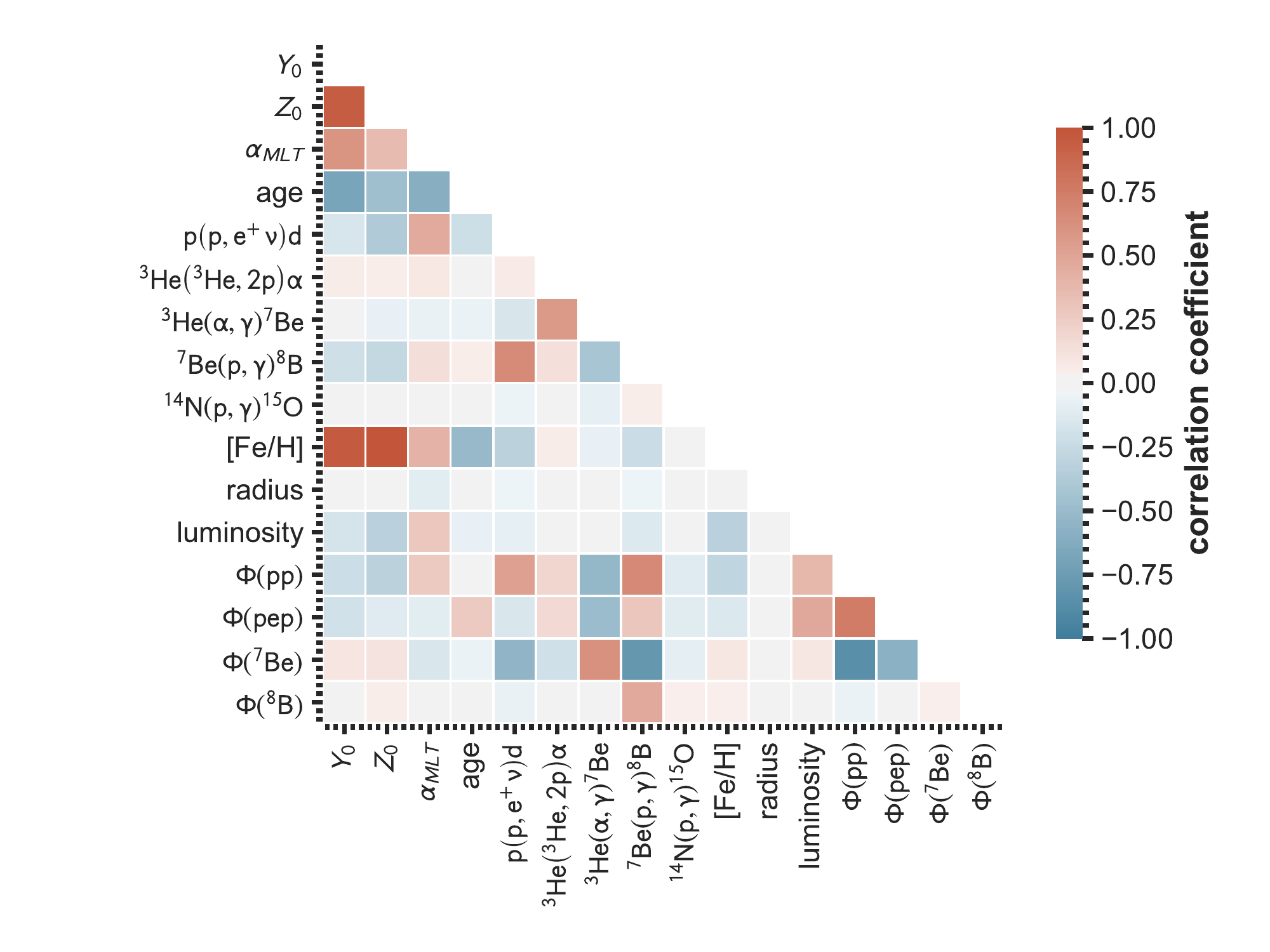}%
    \caption{Correlations between variables in the MCMC analysis of solar data. The first nine are input parameters, and the next seven are outputs of the model. } 
    \label{fig:allcorrs}
\end{figure}

\mb{It is notable that some, but not all, of the nuclear reaction rates measured using solar data shall in fact be highly correlated. 
For example, the correlation coefficient between the $S$-factors of the $\pp{}$ and $\beb{}$ reactions is 0.66. 
\mbb{This is likely due to the fact that increasing the \pp{} rate requires a lower central temperature in order to maintain the luminosity, and hence requires an increase to the \beb{} rate to match the observed $^8$B flux. }
The correlations between other pairs of reactions are shown in Figure~\ref{fig:correlations}. 
\mbb{One can here also see anti-correlation between \hehe{} and \hehef{}, which likely arises due to the opposite effect on the branching ratio  (\emph{cf}.~Figure~\ref{fig:ppchaincno}).}
\mbbbb{The anti-correlation between \beb{} and \hehef{} is also expected in order to maintain the $^8$B neutrino flux.}
The correlations between all the parameters of the model are shown in Figure~\ref{fig:allcorrs}. 
\mbb{One sees here for instance that the mixing length parameter is correlated with the \pp{} rate, and the flux of the pep reaction is correlated with the age of the Sun.}
}

\begin{table}
    \centering
    \caption{Posterior Uncertainties from MCMC Analysis of Solar Data and Factors of Improvement Over Priors Listed in Table~\ref{tab:Sfactors}\label{tab:MCMC}}
    \begin{tabular}{l l l c}\hline
        \multirow{2}{*}{Reaction} & \multicolumn{2}{c}{Uncertainty of $S(0)$} & \multirow{2}{*}{Improvement factor} \\
        & [keV b] & [\%] &  \\ \hline\hline
        \hphantom{$^{14}$}\ppfull{} & $0.023 \times 10^{-22}$ & $0.14\%$ & $7$ \\
        \hphantom{$^{1}$}\tralphiumalpha{} & $0.24 \times 10^{3}$ & $0.89\%$ & $6$ \\
        \hphantom{$^{1}$}\tralphiumbe{} & $0.016$ & $5.2\%$ & $1$ \\
        \hphantom{$^{1}$}\BeB{} & $0.074 \times 10^{-2}$ & $1.7\%$ & $5$ \\
        \NO{} & $0.11$ & $4.2\%$ & $2$ \\
        \hline
    \end{tabular}
\end{table}

\section{Turn-off ages} \label{sec:ages}
Accurate stellar ages are vital for numerous purposes, such as determining the evolution of the galaxy \citep[e.g.,][]{2017AN....338..644M, 2021ApJ...916...88G, 2021arXiv211101669B}, dating exoplanets \citep[e.g.,][]{2015ApJ...799..170C, Bellinger2019}, and providing consistency checks for cosmological theory \citep[e.g.,][]{2021MNRAS.505.2764B}. 
The main-sequence turn-off in particular is an important astrophysical clock for determining the ages of stellar clusters \citep[e.g.,][]{1995ApJ...444L...9C}. 

\begin{figure}
    \centering
    \includegraphics[width=\columnwidth,keepaspectratio]{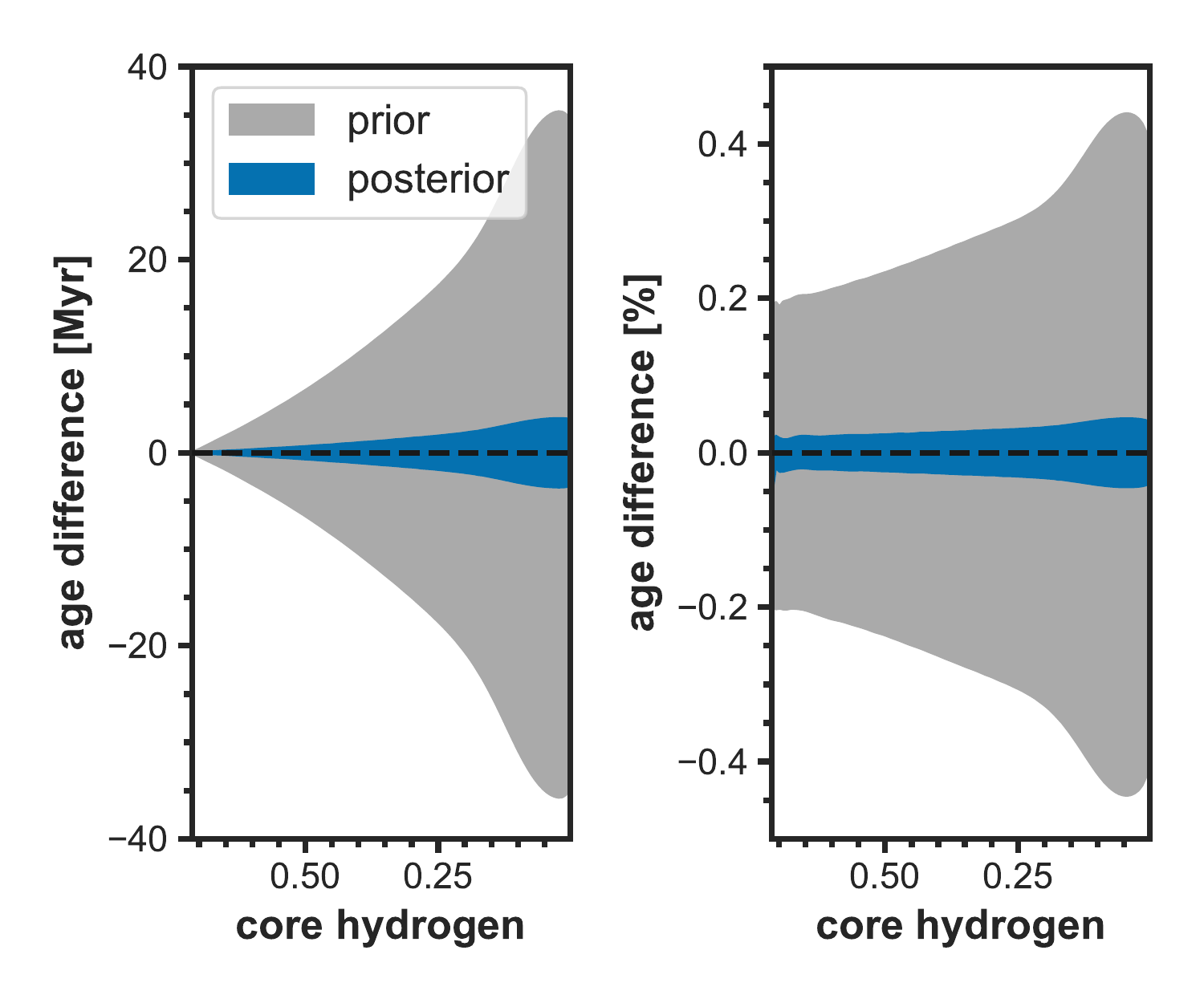}
    \caption{Differences in theoretical ages of a 1~M$_\odot$ star throughout its main sequence evolution \mbb{compared at the same central hydrogen abundance} using current uncertainties on the \pp{} rate (gray shading) and the theoretical minimum uncertainties that are possible from solar data (blue shading). \label{fig:turnoff}}
\end{figure}

Based on the theoretical evolution of the Sun until core hydrogen exhaustion, Figure~\ref{fig:turnoff} compares the age spread at equivalent central hydrogen abundance that one gets with current uncertainties on the \pp{} rate to those that may be possible with solar data. 
The differences in age at the end of the main sequence with current uncertainties span approximately 70~Myr at $1\sigma$, thereby imposing a systematic uncertainty of about 0.44\%. 
The measurements that may be possible from solar data reduce this uncertainty by an order of magnitude, down to 0.044\%. 


We have also investigated the age uncertainties corresponding to the other rates, but found them to be rather negligible in comparison to the \pp{} uncertainty, despite their larger fractional uncertainties. The \hehef{} reaction contributes only 0.14\% uncertainty to the turn-off age of a solar-mass star, and \hehe{} only 0.074\%.

\section{Conclusions} \label{sec:conclusion}
\mb{In this work, we have quantified the constraints that solar measurements are able to place on the rates of nuclear reactions. 
We found that helioseismic data are highly sensitive probes of the \pp{} and \hehe{} reactions. 
Measurements of the $^8$B neutrino flux mainly constrain the possible rate of the \beb{} reaction, but also has some sensitivity to the \pp{} and \hehef{} reactions.}

\mb{Through an MCMC analysis, we combined the helioseismic and neutrino flux data, and found that each of the \pp{}, \hehe{}, and \beb{} reactions can in principle be improved by more than a factor of five using present solar data. 
The analysis showed that solar measurements of the \pp{} and \beb{} rates will be highly correlated, with smaller or no correlations between other pairs of reactions. Future measurements of solar g~modes may further improve the possible constraints on the reaction rates \citep{2021A&A...651A.106S}. }

\section*{Acknowledgements}
We thank Federico~Spada and Selma~de~Mink for useful comments and suggestions that improved the manuscript. 
Funding for the Stellar Astrophysics Centre is provided by The Danish National Research Foundation (Grant agreement no.: DNRF106). 
The numerical results presented in this work were obtained at the Centre for Scientific Computing, Aarhus\footnote{\url{http://phys.au.dk/forskning/cscaa/}}.

\section*{Data Availability Statement}
The data and code underlying this article are available at \url{https://github.com/earlbellinger/solar-reaction-rates}.

\bibliographystyle{mnras}
\bibliography{Bellinger}




\bsp	
\label{lastpage}
\end{document}